\documentstyle[12pt] {article} 
\title{Neutron-Proton High-Energy Charge Exchange Scattering} 
\author{Y. Yan\thanks{e-mail: yupeng@ccs.sut.ac.th}\,\, and R. Tegen\thanks{
Permanent Address: Department of Physics, University of the Witwatersrand, 
P.O. WITS 2050, Johannesburg, South Africa
}
\\
School of Physics, Suranaree University of Technology, \\
Nakhon Ratchasima 30000, Thailand \\
\\
T. Gutsche, V.E. Lyubovitskij and Amand Faessler \\
Institute for Theoretical Physics, T\"ubingen University,\\
Auf der Morgenstelle 14, D-72076 T\"ubingen, Germany}

\begin{document}    
\maketitle
\newpage
\begin{abstract}
The high energy $pn$ charge exchange scattering reaction is studied in 
an effective hadron model for the energy range 
45.9 GeV$^2\le s\le$ 414.61 GeV$^2$. The main features of the observed 
differential cross section, the forward peak and the scaling behavior
over a large energy region, are well reproduced. 
\end{abstract}
\vskip 4cm
---------------   
\newpage
Nucleon-nucleon and nucleon-antinucleon scattering cross sections (total and differential)
have been measured over a very large energy range (from threshold up to several TeV for the
total center-of-mass energy). Such data are the major source of information on the hadronic
and sub-hadronic properties of all matter\cite{expt1}. In this communication we focus
on an interesting sub-set of those data, namely the neutron-proton charge-exchange (CEX)
differential cross sections $d\sigma/dt$ as a function of $s$ which exhibit two main features:
(i) a sharp forward peak which appears for $|t|<0.02$ (GeV/c)$^2$,
and (ii) a scaling behavior, the nearly form-invariant differential cross section
depends on the
lab momentum $p$ in the form of $1/p^2$ over a wide energy region.
These two features represent an enourmous challenge for theoretical nucleon-nucleon
models.
\par
The generally quite successful relativistic meson-exchange models of the nucleon-nucleon
interaction find it very difficult to explain (i) and (ii) without adhoc adjustments to the
underlying model. In any of those meson-exchange models one would expect the lightest
charged mesons ($\pi$ and $\rho$) to dominate the $CEX$ differential cross sections at
least in the low-$|t|$ region. It turns out that the one-pion-exchange alone gives
rise to the observed $1/p^2$ dependence of $d\sigma/dt$ for $|t|>0.1$ GeV/c,
but fails to explain
$d\sigma/dt$ at $t=0$ where this contribution is exactly zero.
\par
It is well known since the 1960's that the rho-exchange does not improve the situation.
In 1963 it was pointed out by Phillips\cite{the1} that the data are consistent with a
nearly constant background field.
Following this original observation, a number of works attempting to
explain the origin of such a background fall into two categories,
namely Regge absorption models (for example, by Henyey $et$ $al.$\cite{the2}
and Kane $et$ $al.$\cite{the3})
and Regge cut models (for example, by Chia\cite{the4}).
Regge absorption models assume that the pion pole is modified by inelastic
channels while Regge cut models evaluate double exchanges.
Those models are with some success in reproducing the $np$ CEX differential
cross section.
\par
More recently the $np$ CEX scattering was studied by Gibbs and
Loiseau\cite{the5} in
a model of one- and two-pion exchanges between the quark constituents of the nucleon.
It is noticed
that in this model the one-pion exchange
cannot give a reasonable result, especially for
the forward peak. With the two-pion exchange included, the model fits
the $np$ CEX diffential cross section quite well. However,
only a very narrow energy region was considered. The P-wave part of the
two-pion exchange is similar to the rho-exchange which
leads to a forward peak. This part, however, is energy independent
for small momentum transfers at high energy and, hence, cannot be expected
to provide a major part of the missing contributions to the nucleon-proton $CEX$
differential cross section over the observed range of energies.
\par
{\it The model} - In this work we study $np$ CEX scattering, based on
that the interaction for the nucleon-nucleon elastic scattering is
mainly mediated by the quark-antiquark sea which might be parametrized
by mesons and a background.
We start with the Lagrangians of the $NN\pi$ and $NN\rho$ coupling
\begin{eqnarray}\label{l1}
{\cal L}_{\pi NN}=g_{\pi NN}\,\overline N \, i\gamma^5 \, \vec\pi \, \vec\tau \, N
\end{eqnarray}
\begin{eqnarray}\label{l2}
{\cal L}_{\rho NN} = g_{\rho NN}\overline N\gamma^{\mu} \, \vec\rho_\mu {\vec\tau}
\,N+\frac{1}{4M_N}\,f_{\rho NN} \overline N\sigma^{\mu\nu} \,
(\partial_\mu\vec\rho_\nu - \partial_\nu\vec\rho_\mu) \, \vec\tau \,N
\end{eqnarray}
with
\begin{eqnarray}
N=
\left(
\begin{array}{c}
p \\
n
\end{array}
\right)
\end{eqnarray}
where $p$, $n$, $\vec\pi$ and $\vec\rho_\mu$ stand
for proton, neutron,
$\pi$ and $\rho$ fields, respectively.
These Lagrangians are widely used in various literatures such as Ref.\cite{Machleidt}.
It is straightforward to calculate $d\sigma/dt$ for high energies
($s$ is much larger
than any mass scale involved) and low $|t|$
with the results
\begin{eqnarray}\label{l3}
\left(\frac{d\sigma}{dt}\right)_\pi\sim t^2/s^2
\end{eqnarray}
and
\begin{eqnarray}\label{l4}
\left(\frac{d\sigma}{dt}\right)_\rho\sim {\rm Constant}
\end{eqnarray}
for one $\pi$ and one
$\rho$ exchanges, respectively.
As discussed above, eq.({\ref{l4}) is inconsistent with the scaling behavior as stated in
(ii), so the one-rho exchange cannot dominate the neutron-proton
$\left(d\sigma/dt\right)_{CEX}$
at large $s$ and low $|t|$.
While giving the correct energy-dependence, eq.$(\ref{l3})$ fails to reproduce the
the forward peak for $\left(d\sigma/dt\right)_{t\sim 0}$.
Obviously there is an additional contribution, beyond $\pi$- and $\rho$-exchange,
necessary to explain the $t$- and $s$-dependence of $d\sigma/dt$.
\par
We study the problem based on the argument that
the nucleon-nucleon elastic scattering is mainly mediated by the
quark-antiquark
sea around the quark core of nucleons, and the quark-antiquark sea might be
parameterised as a four-nucleon contact interaction and by various observed mesons.

We use a model Lagrangian ${\cal L}_{eff}$ which
includes the free part for the nucleons ${\cal L}_N$ and the mesons ${\cal L}_M$,
the meson-nucleon interaction ${\cal L}_{MNN}$ and the four-nucleon
interaction term ${\cal L}_{4N}$ modelling the short range nucleon-nucleon
interactions or the background:
\begin{eqnarray}\label{l5}
{\cal L}_{eff}(x)={\cal L}_{N}(x)+{\cal L}_{M}(x)+
{\cal L}_{MNN}(x)+{\cal L}_{4N}(x),
\end{eqnarray}
where
\begin{eqnarray}\label{l6}
{\cal L}_N(x)=\bar N(x)(i\gamma^\mu\partial_\mu-M_N)N(x),
\end{eqnarray}
\begin{eqnarray}\label{l6_1}
{\cal L}_M(x) = - \frac{1}{2} \vec{\pi}(x) [\Box + M_{\pi}^2] \vec{\pi}(x)
+ \frac{1}{2} \vec{\rho}_\mu(x) [ g^{\mu\nu} (\Box + M_\rho^2)
- \partial^\mu\partial^\nu ] \vec{\rho}_\nu(x),
\end{eqnarray}
\begin{eqnarray}\label{l6_2}
{\cal L}_{MNN}(x) &=&\int d^4y\bar N(x)i\gamma^5 \vec{\pi}(y) \vec{\tau}
G_{\pi NN}(x-y)N(x) \\
&+&\int d^4y\bar N(x)\gamma^\mu\vec{\rho}_\mu(y) \vec{\tau} \,
G_{\rho NN}(x-y)N(x)\nonumber\\
&+&\frac{1}{4M_N}\int d^4y\bar N(x)\sigma^{\mu\nu}
\vec{R}_{\mu\nu}(y) \vec{\tau} \, F_{\rho NN}(x-y)N(x) ,\nonumber
\end{eqnarray}
\begin{eqnarray}\label{l7}
{\cal L}_{4N}(x)=-\frac{g^2}{M_N^2}\int d^4y \bar N(x) N(x) F_{4N}(x-y) \bar N(y) N(y) .
\end{eqnarray}
$\vec{R}_{\mu\nu} = \partial_\nu \vec\rho_\mu - \partial_\mu \vec\rho_\nu$
is the strength of the $\rho$-meson field;
$G_{\pi NN}$, $G_{\rho NN}$ and $F_{\rho NN}$ are form factors modelling
the distribution of the meson cloud in nucleon \cite{FBS96}; $F_{4N}$ is a form factor
modelling the short-ranged nucleon-nucleon interaction or the backgorund.
The Fourier-transform of the vertex form factors is defined as:
\begin{eqnarray}
G(F)_{MNN} (x) = \int d^4p \, e^{-ipx} \, G(F)_{MNN}(p^2)
\end{eqnarray}
\begin{eqnarray}
F_{4N} (x) = \int d^4p \, e^{-ipx} \, F_{4N}(p^2)
\end{eqnarray}
The $\pi NN$ coupling constant is fixed from the Goldberger-Treiman relation:
\begin{eqnarray}
G_{\pi NN}(m_\pi^2) = g_A \frac{M_N}{F_\pi}
\end{eqnarray}
where $g_A = 1.267$, $F_\pi = 93$ MeV and $M_N=938$ MeV are the experimental
values of axial nucleon charge, weak pion decay constant and nucleon mass
resulting in
\begin{eqnarray}
\frac{G^2_{\pi NN}(m_\pi^2)}{4\pi}=13.0
\end{eqnarray}
For the $\rho NN$ coupling constants we use the values fixed
in low and medium energy $NN$ reactions \cite{Machleidt}
\begin{eqnarray}
\frac{G^2_{\rho NN}(m_\rho^2)}{4\pi}=0.84,\,\,\,\,\,
\frac{F_{\rho NN}(m_\rho^2)}{G_{\rho NN}(m_\rho^2)}=6.1 .
\end{eqnarray}

The background field contribution, eq.(\ref{l7}), is reminiscent of
the Nambo-Jona-Lasinio (NJL) 4-fermion interaction\cite{njl1} which in the
chain approximation (RPA) gives rise to a pionic mode and a scalar quark-antiquark
condensate. Here we adjust the 4-fermion coupling strength $g$ in eq.(\ref{l7}) to
the experimental $np$ data. Due to the analogy between eq.(\ref{l7}) and the NJL
Lagrangian we will consider the background contribution to have similar properties
as if a '$\sigma$'-like object was exchanged between two nucleons, hence
we will adopt a non-trivial vertex function.

The quark-antiquark substructure of $N$, $\pi$, and $\rho$ is assumed to be manifest
in the non-trivial meson-nucleon vertex functions.
We study various forms of those vertex functions such as monopole,
dipole, multipole and exponential forms.
We find that the experimental
data strongly suggest the monopole form for the $\pi NN$ vertex function,
and favor
the tripole form for the $\rho NN$ vertex function.
For the background contribution, we adopt a dipole
form.
The t-dependent vertex functions are defined as
\begin{eqnarray}
F_{4N}(t)=\frac{1}{(1-t/\Lambda^2)^2},
\end{eqnarray}
\begin{eqnarray}
G_{\pi NN}(t)=G_{\pi NN}(m^2_\pi)\frac{1-m^2_\pi/\Lambda_\pi^2}{1-t/\Lambda_\pi^2}
\end{eqnarray}
and
\begin{eqnarray}
G(F)_{\rho NN}(t)=G(F)_{\rho NN}(m^2_\rho)=
\left(\frac{1-m^2_\rho/\Lambda_\rho^2}{1-t/\Lambda_\rho^2}\right)^3 .
\end{eqnarray}
Here, in addition to the coupling strength $g$, the cutoffs $\Lambda$, $\Lambda_\pi$ and
$\Lambda_\rho$ are free parameters adjusted to the experimental data.

In Fig. 1 we indicate the theoretical results with the dashed lines for
the one $\pi$ exchange plus the background contribution (Model A) and
the solid lines for the one $\pi$ and one $\rho$ exchanges plus the
background contribution (Model B).
All the relevant parameters are listed in Table 1 both for Model A and B.
Here $g(f)_M \doteq G(F)_{MNN}(M^2)$ are input parameters while
other parameters are adjusted to the observed differential cross sections
of the $pn$ charge exchange scattering. The energy range considered here
is 45.9 GeV$^2\leq s\leq 414.61$ GeV$^2$.
\begin{table}
\caption{Coupling constants and cutoff parameters employed in Model A and B.}
\begin{center}
\begin{tabular}{|c|c|c|c|c|c|c|c|}
\hline
Models & $g_\pi^2/4\pi$ & $g_\rho^2/4\pi$ & $f_\rho/g_\rho$ & $g^2/4\pi$ &
$\Lambda$ [GeV] & $\Lambda_\pi$ [GeV] & $\Lambda_\rho$ [GeV] \\
\hline
Model A & 13.0 & 0.84 & 6.1 & 5.5 & 0.26 & 0.52 & -- \\
\hline
Model B & 13.0 & 0.84 & 6.1 & 5.0 & 0.24 & 0.50 & 1.0 \\
\hline
\end{tabular}
\end{center}
\end{table}
As apparent from Fig.1, the experimental data are well reproduced by Model B.
Model A (one $\pi$ exchange plus the background contribution) reproduces
the data well up to $s=109.68$ GeV$^2$, and the $\rho$ exchange
contribution becomes significant when $s$ is larger than 120 GeV$^2$.

The $np$ CEX differential cross section is well reproduced in a simple model of
one $\pi$ and one $\rho$ exchanges including an additional background contribution.
The background contribution is important only for very small $|t|$.
The one $\rho$ exchange contribution is negligible below $s=100$ GeV$^2$,
but becomes important for $s$ larger than 200 GeV$^2$, particularly
for small $|t|$. The one $\pi$ exchange is important for all energies,
in particular for large $|t|$.

\begin{description}
\item
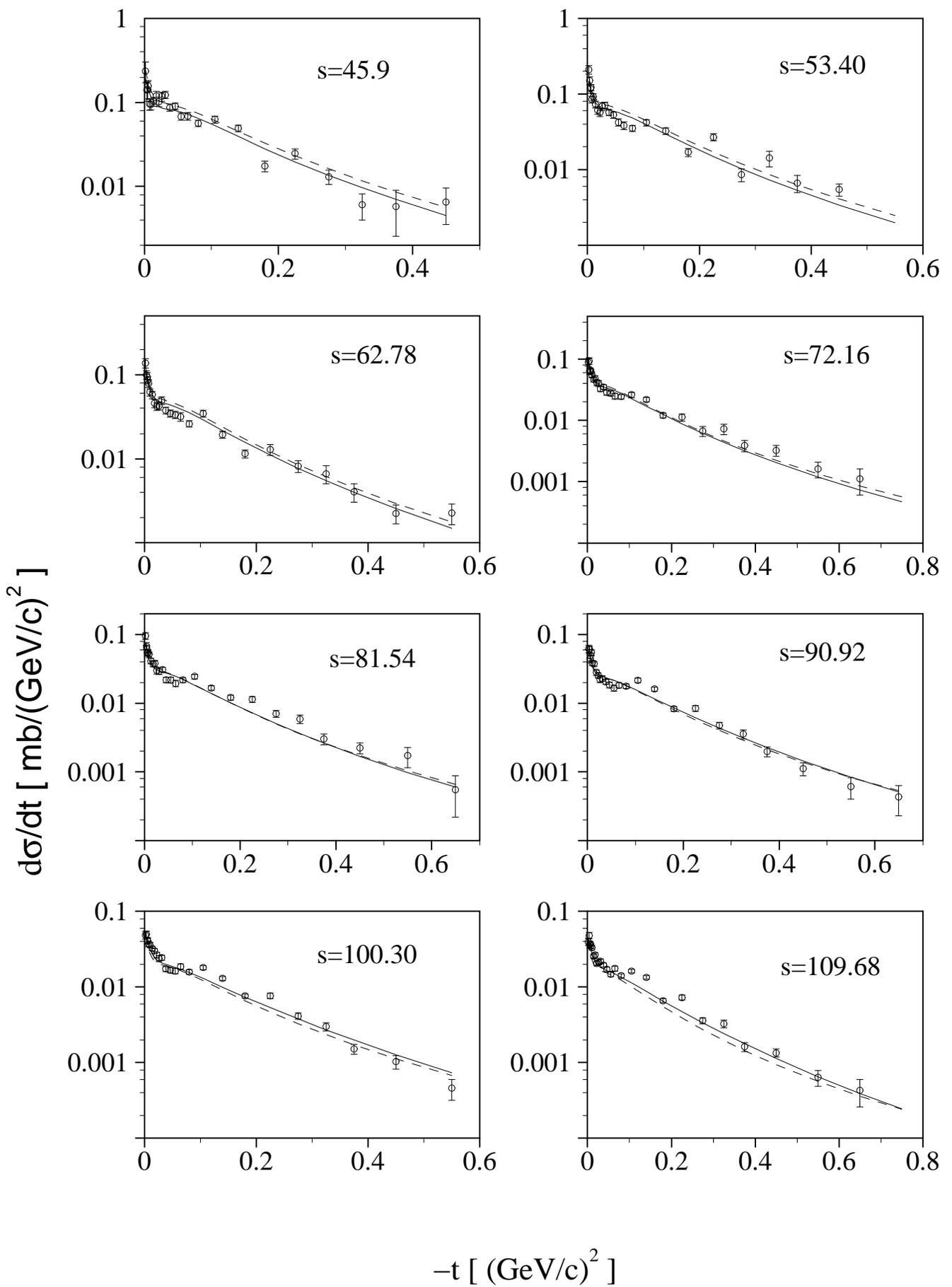
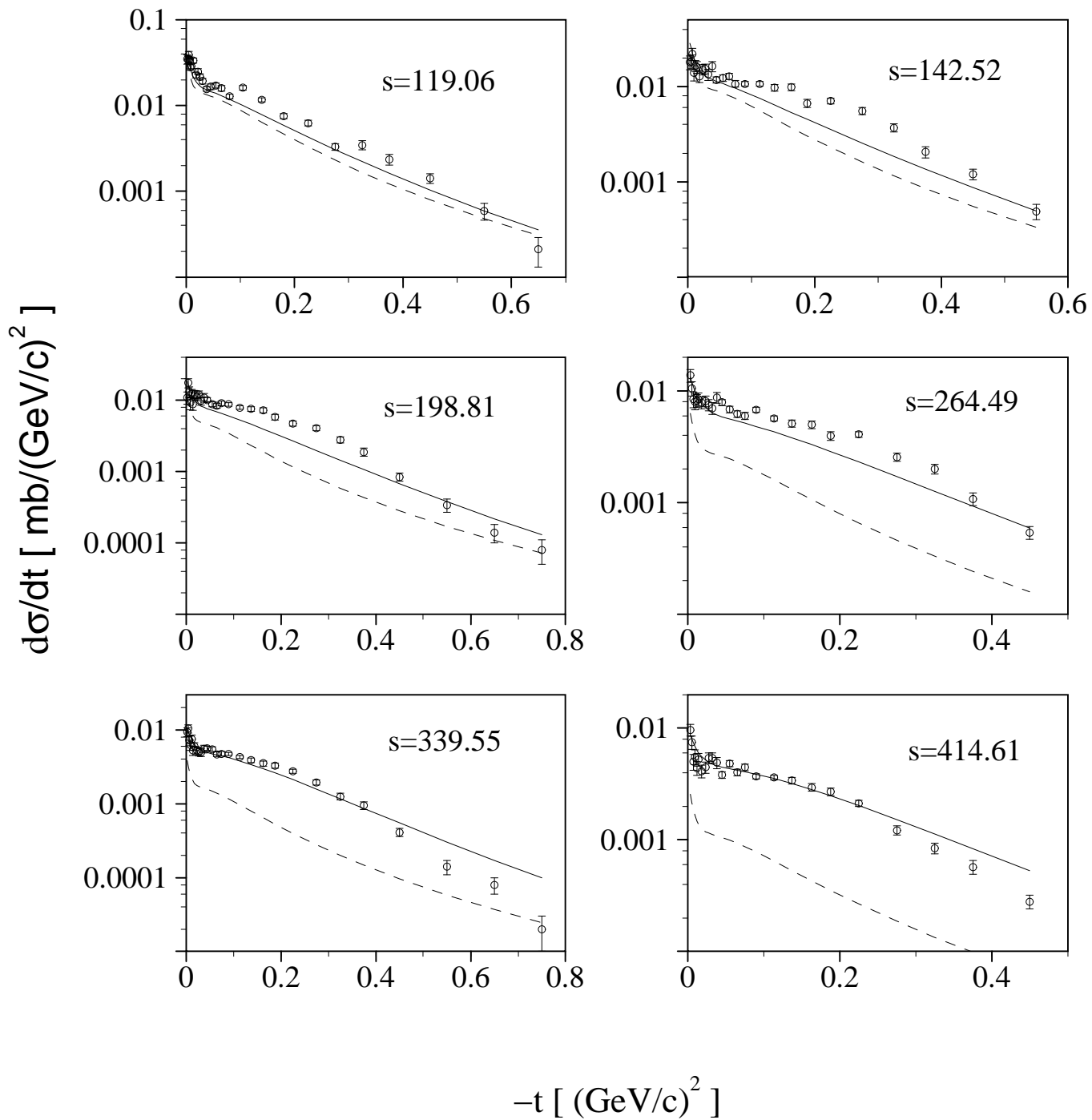
[Fig. 1] Theoretical predictions for $d\sigma^{CEX}/dt$ compared to
experimental data
(taken from the compilation of Ref. \cite{expt1}).
Dashed curve for Model A (one $\pi$-exchange plus background contribution);
solid curve for Model B (one $\pi$ and one $\rho$ exchanges plus background
contribution). Here $s$ are in GeV$^2$.
\end{description}

\end{document}